\documentclass[a4paper,fleqn,usenatbib]{mnras}

\usepackage[T1]{fontenc}
\usepackage{ae,aecompl}

\usepackage{graphicx}	
\usepackage{amsmath}	
\usepackage{amssymb}	
\usepackage{color}
\usepackage{array,multirow} 
\usepackage{makecell} 
\newcommand{\aref}[1]{\hyperref[#1]{Appendix~\ref{#1}}}

\newcommand{\xamin}{\texttt{XAmin }}

\title[CNN classification of galaxy cluster candidates]{Multiwavelength classification of X-ray selected galaxy cluster candidates using convolutional neural networks}

\author[Kosiba et al.]{Matej Kosiba,$^{1, 2}$\thanks{E-mail: matej.kosiba@gmail.com}
Maggie Lieu,$^{2,3}$ 
Bruno Altieri,$^{2}$ 
Nicolas Clerc,$^{4}$ 
Lorenzo Faccioli,$^{5}$ 
\newauthor
Sarah Kendrew,$^{6}$ 
Ivan Valtchanov,$^{7}$
Tatyana Sadibekova,$^{5,8}$
Marguerite Pierre,$^{5}$
\newauthor
Filip Hroch, $^{1}$
Norbert Werner, $^{9,1,10}$
Luk\'{a}\v{s} Burget,$^{11}$
Christian Garrel, $^{5}$
\newauthor
Elias Koulouridis,$^{12,5}$
Evelina Gaynullina,$^{8}$
Mona Molham,$^{13}$
Miriam E. Ramos-Ceja$^{14}$
\newauthor
and Alina Khalikova$^{8}$\\
$^{1}$Department of Theoretical Physics and Astrophysics, Faculty of Science, Masaryk University, Kotl\'a\v rsk\'a 2, Brno, 611 37, Czech \\
Republic\\
$^{2}$European Space Astronomy Centre, ESA, Villanueva de la Ca$\tilde{n}$ada, E-28691 Madrid, Spain\\
$^{3}$Centre for Astronomy and Particle Theory, University of Nottingham, UK\\
$^{4}$IRAP, Universit\'e de Toulouse, CNRS, CNES, UPS, (Toulouse), France\\
$^{5}$AIM, CEA, CNRS, Universit\'e Paris-Saclay, Universit\'e Paris Diderot, Sorbonne Paris Cite, F-91191 Gif-sur-Yvette, France\\
$^{6}$European Space Agency, Space Telescope Science Institute, 3700 San Martin Drive, Baltimore MD 21218, USA\\
$^{7}$Telespazio Vega UK for ESA, European Space Astronomy Centre, Operations Department, 28691 Villanueva de la Ca\~nada, Spain\\
$^{8}$Ulugh Beg Astronomical Institute of Uzbekistan Academy of Science, 33 Astronomicheskaya str., Tashkent, UZ-100052, Uzbekistan\\
$^{9}$MTA-E\"otv\"os University Lend\"ulet Hot Universe Research Group, P\'azm\'any P\'eter s\'et\'any 1/A, Budapest, 1117, Hungary\\
$^{10}$School of Science, Hiroshima University, 1-3-1 Kagamiyama, Higashi-Hiroshima 739-8526, Japan\\
$^{11}$Faculty of Information Technology, Brno University of Technology, Bo\v{z}et\v{e}chova 2, Brno, 612 00, Czech Republic \\
$^{12}$Institute for Astronomy \& Astrophysics, Space Applications \& Remote Sensing, National Observatory of Athens, GR-15236 Palaia \\
Penteli, Greece\\
$^{13}$National Research Institute of Astronomy and Geophysics (NRIAG), 11421 Helwan, Egypt\\
$^{14}$Max-Planck Institut f\"ur extraterrestrische Physik, Postfach 1312, 85741 Garching bei M\"unchen, Germany\\
}

\date{Accepted XXX. Received YYY; in original form ZZZ}

\pubyear{2020}

\begin{document}
\label{firstpage}
\pagerange{\pageref{firstpage}--\pageref{lastpage}}
\maketitle

\begin{abstract}
Galaxy clusters appear as extended sources in \textit{XMM-Newton} images, but not all extended sources are clusters. So, their proper classification requires visual inspection with optical images, which is a slow process with biases that are almost impossible to model. We tackle this problem with a novel approach, using convolutional neural networks (CNNs), a state-of-the-art image classification tool, for automatic classification of galaxy cluster candidates. We train the networks on combined \textit{XMM-Newton} X-ray observations with their optical counterparts from the all-sky Digitized Sky Survey. Our data set originates from the X-CLASS survey sample of galaxy cluster candidates, selected by a specially developed pipeline, the \texttt{XAmin}, tailored for extended source detection and characterisation. Our data set contains 1\,707 galaxy cluster candidates classified by experts. Additionally, we create an official Zooniverse citizen science project, \textit{The Hunt for Galaxy Clusters}, to probe whether citizen volunteers could help in a challenging task of galaxy cluster visual confirmation. The project contained 1\,600 galaxy cluster candidates in total of which 404 overlap with the expert's sample. The networks were trained on expert and Zooniverse data separately. The CNN test sample contains 85 spectroscopically confirmed clusters and 85 non-clusters that appear in both data sets. Our custom network achieved the best performance in the binary classification of clusters and non-clusters, acquiring accuracy of 90 \%, averaged after 10 runs. The results of using CNNs on combined X-ray and optical data for galaxy cluster candidate classification are encouraging and there is a lot of potential for future usage and improvements.

 \end{abstract}




\section{Introduction}
Galaxy clusters are massive systems at the peaks of the cosmic web. Their composition, rich in dark matter and hot baryonic gas makes them a potentially powerful tool to constrain cosmological parameters, growth of structure, neutrino mass and sterile neutrinos through cluster number counts, the cluster mass function and the baryon fraction \citep{Allen2011, Mantz2015, Bohringer2016}. 

In recent years, large cluster surveys such as XXL \citep{Pierre2016,Pacaud2016}, XCS \citep{Mehrtens2012}, X-CLASS \citep{Clerc2012b,Ridl2017}, \textit{Planck} \citep{Bartlett2008}, redMaPPer \citep{Rykoff2014}, or the SPT-SZ survey \citep{Bleem2015} have made it possible to statistically improve constraints on cosmology. However one of the challenges in using galaxy clusters for cosmology is understanding and modelling of the cluster selection function \citep[e.g.][]{Pacaud2006}. The selection function has to be modelled in terms of observable parameters (like flux and apparent size), which can then be converted into galaxy cluster mass for a given cosmology and galaxy cluster physics evolution. The selection function of galaxy clusters is not trivial to model and often oversimplified.
A selection function should not only take into account the volume and redshift of the survey but also the choice of clusters, which is often more complicated than a cut in flux. In X-ray wavelengths, 
whilst extended emission is generally a robust indicator of a galaxy cluster, the emission can also be attributed to nearby galaxies, saturated AGN and unresolved double point-sources. For this reason, galaxy cluster candidates are still visually examined together with optical data, prior to any spectroscopic confirmation \citep{Adami2018}. This process is tedious and out-dated with uncertainties impossible to model. With large X-ray sky surveys such as \textit{e-ROSITA} \citep{Merloni2012} expecting to discover tens of thousands of new galaxy clusters, combined with large optical surveys including LSST \citep{Ivezic2008} and \textit{EUCLID} \citep{Racca2016}, the old techniques will become obsolete. We need to prepare for the future with new methods that are able to deal with big data and improved accuracy.

Citizen science projects proved to be a great asset for scientific problems where human classifications are required for large amounts of data \citep[e.g.][]{Lintott2008, Willett2013}. In the first version of the most well known of all citizen science projects, the Galaxy Zoo \citep{Lintott2008}, citizen volunteers managed to achieve more than 90\,\% agreement with experts in a task of morphological classification of galaxies. While citizen projects are intended to provide huge manpower in the assessment of large astronomical data sets, the question whether this is an advantage over a limited number of evaluations by experts in the case of the confirmation of galaxy cluster candidates remains to be addressed. This paper scrutinizes this issue by evaluating the citizen volunteers success rate.

Machine learning offers a more constructive approach to the problem. The power of Machine learning has been demonstrated in astronomy for more than two decades, with applications including star-galaxy discrimination \citep{Odewahn1992, Bertin1993}, classification of galaxy spectra \citep{Folkes1996}, photometric redshift estimation \citep{Collister2004} or anomaly detection in X-ray spectra \citep{Ichinohe2019}, to name a few. With the introduction of Convolutional Neural Networks \citep[CNNs,][]{LeCun1999} and deep learning \citep{Hinton2007}, it has been possible to automate human vision tasks such as image recognition \citep[see e.g.][]{Goodfellow2014, Schawinski2017, Ackermann2018, Lieu2018}. 

Supervised learning with convolutional neural networks (CNNs) was designed specifically for image classification tasks. If the true labels (classification classes) of the images are known, they can be used to train CNNs.
The current way galaxy clusters are classified are liable to false positives and false negatives. Galaxy cluster candidates picked by an automated pipeline are visually analysed by several experts to create an initial catalogue of galaxy clusters, that are later verified with a spectroscopic confirmation. This process will not scale with large data volumes. Citizen science allows us to harness a large number of opinions on each object classification on a short timescale, speeding up the process significantly yet having a reasonable agreement with experts \citep[see e.g.][]{Willett2013, Dieleman2015}.
CNNs can be then trained on classifications made by either experts or citizen volunteers or both, to automate the final classification of galaxy cluster candidates, or even skipping the first step of the pipeline picking the candidate clusters. Applying CNN selection on simulations will enable modelling the selection function.

In this paper, we introduce a citizen science project we created to obtain large numbers of classified objects. We compare the performance of citizen volunteers with experts. We train CNNs on classifications of citizen volunteers and experts and compare their results. CNNs are tested on spectroscopically confirmed galaxy clusters and objects classified as non-clusters by experts.

The structure of the paper is as follows: in \hyperref[sec:The Hunt for Galaxy Clusters]{Section~2} we present our citizen science project and its development together with a description of the observations and the construction of their classifications by the experts,  in \hyperref[sec:MachineLearningApproach]{Section~3} we introduce the machine learning methods we use,
\hyperref[performance_measurements]{Section~4} presents measurements used to evaluate classification or detection performance,
\hyperref[sec:Results and Discussion]{Section~5} presents the results of the citizen science campaign as well as the results and discussion of neural networks analysis. Finally, we conclude in \hyperref[sec:Conclusions]{Section~6}.

\section{The Hunt for Galaxy Clusters}
\label{sec:The Hunt for Galaxy Clusters}

Our citizen science project, \textit{The Hunt for Galaxy Clusters}\footnote{https://www.zooniverse.org/projects/matej-dot-kosiba/the-hunt-for-galaxy-clusters}, was launched online as an official Zooniverse project on the 24th of October 2018. There were 1\,600 galaxy cluster candidates in the project that have been detected as extended X-ray sources by the \xamin\ wavelet-based pipeline \citep{Pacaud2006}. Each object was classified by at least 30 different volunteers and this was completed by the 29th of April 2019. 1\,227 volunteers participated in the project. Classifications of not logged in volunteers, as well as classifications which have been done on each object multiple times by the same volunteer, were not considered. 

\begin{figure}
    \begin{center}
    \includegraphics[width=0.23\textwidth]{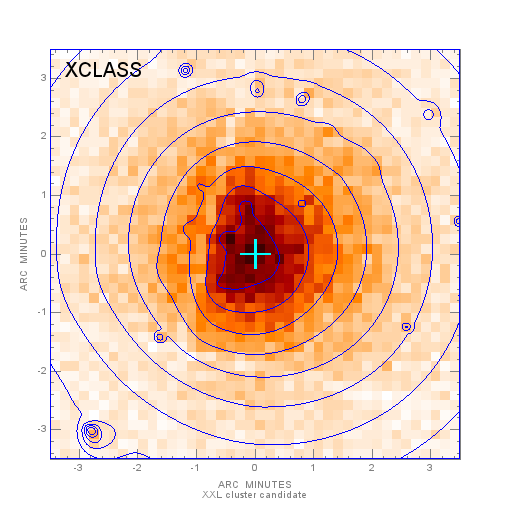} \hfill
    \includegraphics[width=0.23\textwidth]{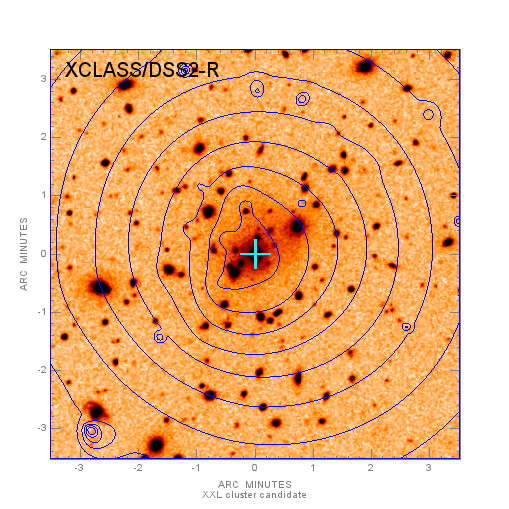}
    \end{center}
    \begin{center}
    \includegraphics[width=0.23\textwidth]{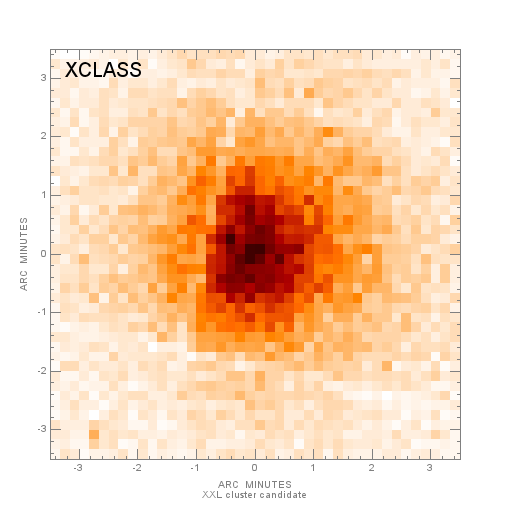} \hfill
    \includegraphics[width=0.23\textwidth]{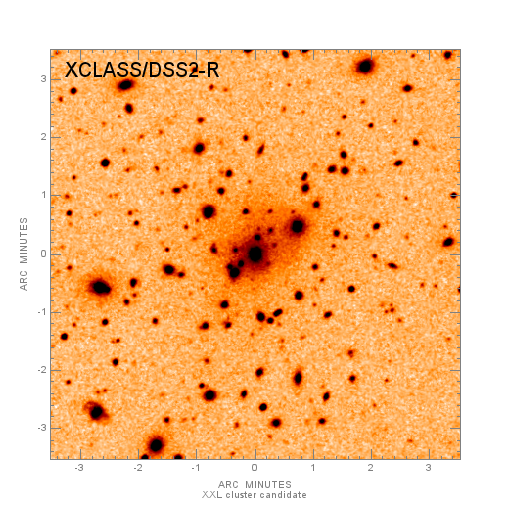}
    \end{center}
    \caption{\textit{Top left}: raw X-ray image with contours showing the areas of constant X-ray brightness and a cyan cross marking the object selected for classification. \textit{Bottom left}: raw X-ray image without contours and markings. \textit{Right}: corresponding optical images.} 
    \label{fig:data_format_in_zooniverse}
\end{figure}

The project starts with a short tutorial briefly explaining how to navigate in the project's page and how to classify candidate clusters. Each object comes with four images, covering the exact same area of the sky (7$\times$7 arcmin$^2$): two X-ray and two optical images. \autoref{fig:data_format_in_zooniverse} shows all four images of a galaxy cluster candidate as shown to the volunteers in \textit{The Hunt for Galaxy Clusters}.

\begin{figure}
	\includegraphics[width=\columnwidth]{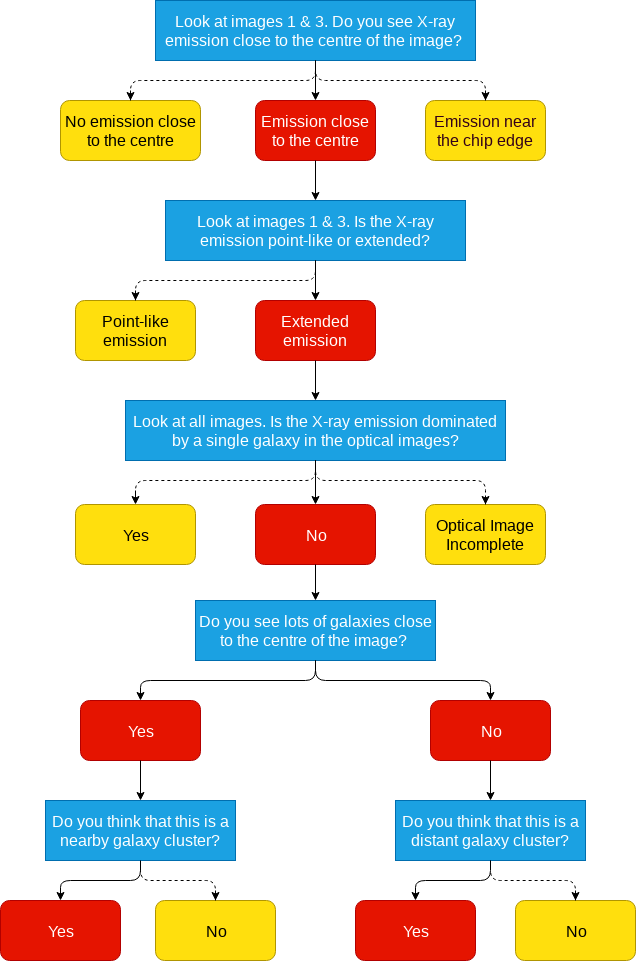}
	\caption{The decision tree of \textit{The Hunt for Galaxy Clusters} Zooniverse citizen science project. Blue cells represent questions, red are answers leading to the \textit{cluster} class and yellow are answers leading to the \textit{non-cluster} class.}
	\label{fig:decision_tree_new}
\end{figure}

Our project uses six questions to help determine the class of a galaxy cluster candidate. Each question has two or three possible answers, and due to the structure of the decision tree (\autoref{fig:decision_tree_new}), only a subset of the questions are answered. Those questions come with help notes, example images, as well as descriptions to each answer. We selected example images very carefully to cover a broad range of objects and/or instrument effects,
in order to avoid biases. The Zooniverse volunteer's answers were then used to create a binary classification scheme of \textit{cluster} and \textit{non-cluster}.

\subsection{Data}
\label{subsec:Data}
The data in this work originates from the XMM CLuster Archive Super Survey (X-CLASS) \citep{Clerc2012}, an X-ray galaxy cluster search in the archival data of the \textit{European Space Agency's} X-ray observatory \textit{XMM-Newton}, combined with corresponding optical counterparts from the Digitized Sky Survey POSS-II (DSS2). We used \textit{XMM-Newton} data obtained between 2000 and 2015, employing selection criteria described in \citep{Clerc2012}, and excluding the data used by the XXL survey \citep{Pierre2016}.  

\subsection{X-ray pipeline}
\label{X-ray pipeline}


Our sample of galaxy cluster candidates has been constructed using the intermediate \xamin 3.5 version (new source models added: double point-source and point\,$+$\,extended source). This version, after the processing of the X-CLASS survey, appeared to suffer from a miss-centering problem randomly affecting a tiny fraction of the point-source population, that led to classify them as extended.  In order to remove miss-classified sources, experts then performed an in-depth screening of the putative cluster candidate lists. The screening dealt as well with usual nearby galaxies and saturated AGNs, that both appear extended in the X-ray images

The pipeline is briefly described below. Firstly, a combined MOS1$+$MOS2$+$PN image of an \textit{XMM-Newton} \citep{Jansen1999} observation is smoothed with a dedicated wavelet smoothing program called mr$\_$filter, described by \cite{Starck1998a} and shown in \cite{Starck1998b} to effectively
recover structures in X-ray images characterised by low numbers of photons.

Secondly, the wavelet smoothed image is analysed by the source extraction
software \texttt{SExtractor} \citep{Bertin1996}. It creates a list
of candidate sources for further analysis, returning an estimate of their
position and their flux.

Note that, since \texttt{SExtractor} was developed for optical images which contain many more photons than the X-ray ones, smoothing the X-ray image is a necessity as \texttt{SExtractor} would not be able to work with raw data. This smoothing can be performed in several ways; the wavelet smoothing used by \xamin is one
of the possible ways of smoothing the image and was shown by \cite{Valtchanov2001}
to give the best results for X-ray images of diffuse sources like galaxy clusters.

Finally, we characterise the candidate sources found by \texttt{SExtractor}. This is done by fitting both a point source model given the \textit{XMM-Newton}
PSF computed at the source position and an extended $\beta$ model \citep{Cavaliere1976}
which better describes galaxy clusters. A source is declared to be a point source
(AGN or an extended source too faint to be characterised as extended) or an extended source (galaxy cluster) depending on which of these two models best fits the
candidates source. The details, including the relevant formulas and the selection criteria
for defining an (almost) pure sample of galaxy clusters, are given in
\cite{Pacaud2006}. 

Coordinates of the galaxy cluster candidates picked by \xamin are then used to produce normalised images (\aref{Appendix:ImPreprocessing}) with and without X-ray contours to show lines of constant X-ray brightness. These contours are superimposed onto the optical counterpart image, together with a cyan cross mark and are used only for human screening to help visualise the X-ray emission.

\subsection{Weighting volunteers classifications}
\label{sec_weighting_volunteers_classifications}
 
Since each object is classified by 30 volunteers, we may end up with different classifications for the same galaxy cluster candidate. Each person's classification ability may vary according to the class and the question asked, and there may even be volunteers who purposely create malicious classifications. To mitigate those effects, we weight classifications of each user question-wise. Weighting is done according to the agreement of the majority, so each user has an accuracy determining a portion of his/her classifications being in agreement with the majority of votes, which is done question-wise,
\begin{align}
& G_i = \frac{C_i}{Q_i}, & i \in {1,...,6}
\label{weighting_accuracy}
\end{align}
where $G_i$ is the weight applied for an individual on question $i$, $C_i$ is the number of answers to question $i$ given by the individual that were in agreement with the majority and $Q_i$ is the total number of answers the individual has made for question $i$. $G_i$ essentially describes the ability of an individual to classify as the majority of volunteers would. Every classification in the project is then weighted according to the classifying volunteer's accuracy for the specific question. The bottom red leafs of the decision tree (\autoref{fig:decision_tree_new}) are classification ending answers corresponds to the final answers stating that the classified object is a galaxy cluster. Similarly, all yellow leafs corresponds to the final answers stating that the object is not a galaxy cluster. Each galaxy cluster candidate gets 30 votes, each vote is an accuracy of the voting user for the question of his/her classification ending answer (one of bottom red leafs or any yellow leaf). Those 30 weighted scores are summed to galaxy cluster (bottom red leafs) and non-galaxy cluster (yellow leafs) categories. The higher score determines the final Zooniverse weighted classification for the galaxy cluster candidate.

\subsection{Classifications of experts}
\label{subsec:Classifications_of_experts}

\begin{figure}
	\centering
	\includegraphics[width=0.23\textwidth]{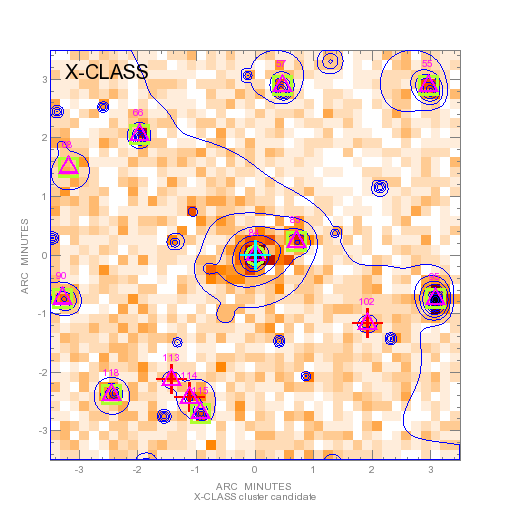} \hfill
	\includegraphics[width=0.23\textwidth]{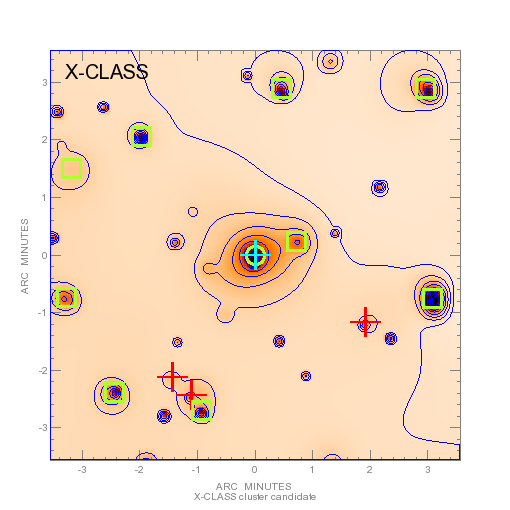} 
	
	\includegraphics[width=0.23\textwidth]{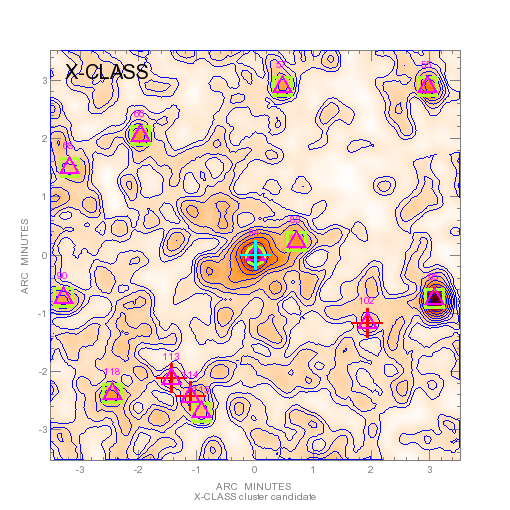}
	\hfill 
	\includegraphics[width=0.23\textwidth]{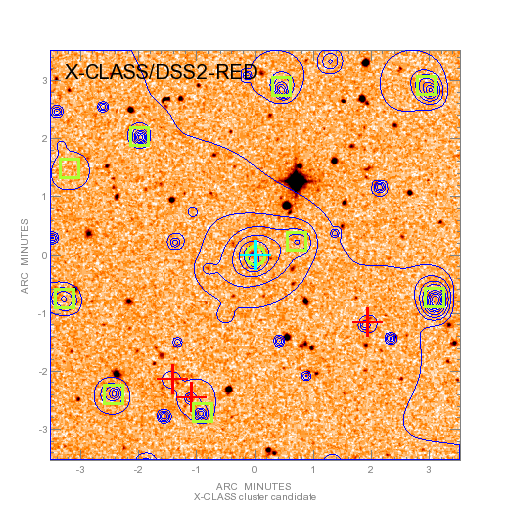}
	
	\includegraphics[width=0.33\textwidth]{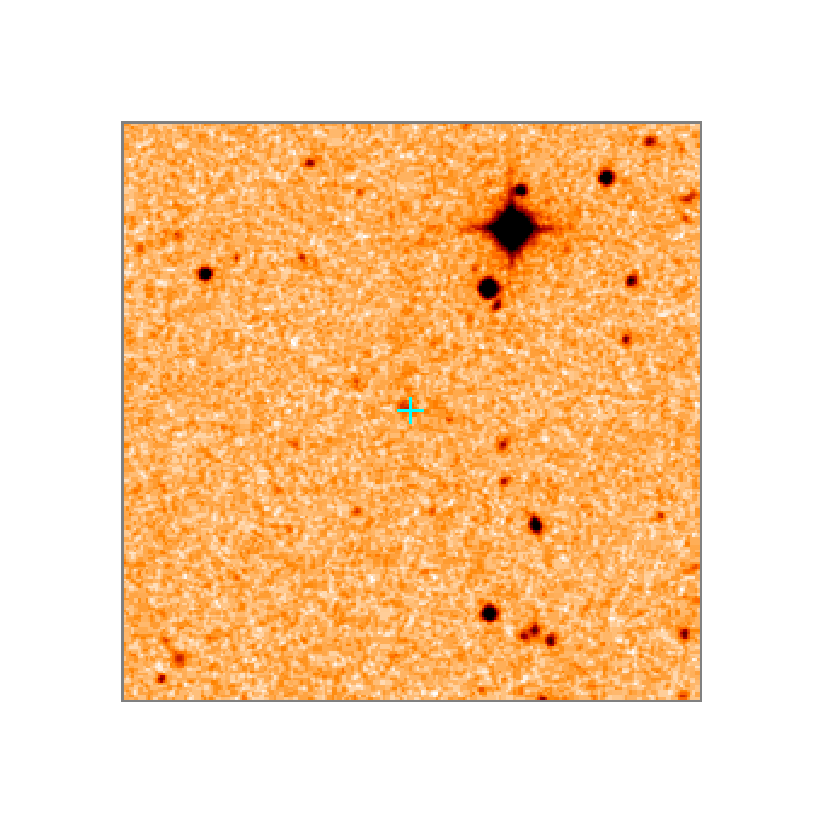}
	\caption{Images of a galaxy cluster candidate classified by experts. Top left: an X-ray raw image overplotted with contours showing areas of constant X-ray brightness, and marks produced by the \xamin\ pipeline. Top right and bottom left images are smoothed versions of the X-ray images, wavelet and Gaussian smoothing produced by the \xamin pipeline, respectively. The Gaussian smoothed image is overplotted with Gaussian contours, the sigma is chosen to be 3 pixels (with a pixel size of 2.5 arc seconds so the sigma is 7.5 arc seconds). Bottom right: the optical counterpart of the X-ray image with superimposed marks and wavelet X-ray contours. All images cover the exact same area of the sky, 7$\times$7 arcmin$^2$, except for the bottom panel, where we focus in the central region (4$\times$4 arcmin$^2$) of the optical image, because with the contours and the symbols it is not easy to see the central cluster brightest galaxy and overdensity of faint galaxies.}
	\label{lotto_example_images}
\end{figure}

The galaxy cluster candidates generated by the \xamin\ pipeline are manually classified by the X-CLASS collaboration. Each galaxy cluster candidate is classified by two experts and three moderators make the final classification on conflicting decisions. \autoref{lotto_example_images} shows how a galaxy cluster candidate is presented to the experts. The images are provided without redshift or sky coordinate information, and the experts make decisions without consulting with each other to avoid any bias. The experts were given the opportunity to classify objects as a low redshift cluster ($0 < z < 0.3$), high redshift cluster ($z > 0.3$), nearby galaxy, point source, star or AGN, double source, artefact, edge, fossil group, high background image, no optical image or dubious source. We create a binary classification scheme where the last four categories in the list are not used, low and high redshift clusters are collectively referred to as clusters and the remaining classes are collectively referred to as non-clusters.

\section{Machine learning approach }
\label{sec:MachineLearningApproach}

Now, we turn our attention to a machine learning approach, which allows us to automatically process astronomical data on much larger scales than what is possible to achieve by human annotations. We use neural networks -- a parametric model, that is able to learn to approximate a complex function from training examples of inputs and the corresponding outputs. In our case, each training example consists of combined X-ray and optical image as the input and the corresponding output class label obtained from a human annotator. In our experiments, we consider binary classification, where the class labels are \textit{galaxy cluster} and \textit{non-cluster}, but also multi-class classification with subcategories that will be discussed in Section~\ref{Multi-class classification}. From the training examples, our neural networks learn to predict posterior probabilities of all classes given an input image. In our experiments, we evaluate the performance of the neural networks using measures discussed in Section~\ref{performance_measurements}. For some of the measures, we need to make a hard classification decision for each input image from our evaluation set. In such a case, we simply select the most probable class. 

In this work, we use Convolutional Neural Networks (CNN), which is currently the most popular and very effective neural network architecture for image processing \citep{LeCun1989,Ciresan2102,Krizhevsky2012}. A deeper knowledge of CNNs is not necessary for interpreting our results and understanding the presented analyses. It is only necessary for understanding some of the technical details. This paper also can not give a complete tutorial to CNNs, therefore, we do not provide a further introduction to CNNs and we kindly refer the interested reader to the relevant textbooks~\citep{Goodfellow-et-al-2016,Bishop2006} or the numerous tutorials available online. We use two CNNs architectures for our experiments: Using the Keras toolkit~\citep{chollet2015}, we build and train our \textit{custom network}, which uses a conventional CNN architecture with interleaving convolutional and pooling layers and final dense layers. The second architecture is MobileNet~\citep{Howard2017}. We take these networks as provided by its authors pre-trained on the ImageNet~\citep{Deng2009} data, which is a large data set of millions of real-word images categorised into thousands of classes. We assume that such pre-training can serve as a good initialisation of the CNN parameters, which are further retrained on our training data for galaxy cluster classification.

\subsection{Data preprocessing}
\label{Data preprocessing}

\begin{figure}
    \centering
    \includegraphics[width=0.15\textwidth]{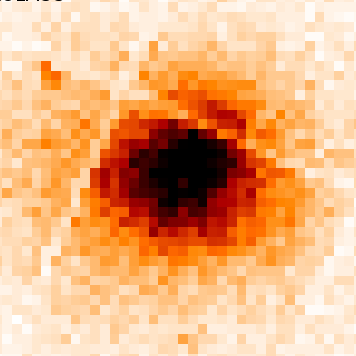} \hfill
    \includegraphics[width=0.15\textwidth]{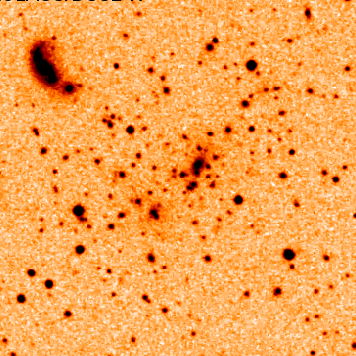} \hfill 
    \includegraphics[width=0.15\textwidth]{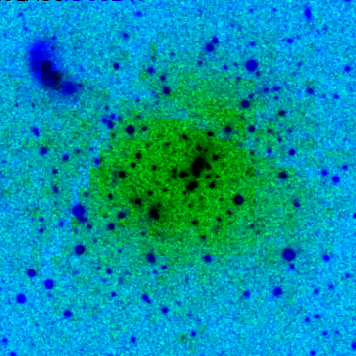}
    \caption{Left is the 356$\times$356 pixel X-ray rgb .PNG image, middle is its 356$\times$356 pixel optical .PNG counterpart and right is an rgb .PNG image made by stacking grayscaled optical image as blue channel, grayscaled X-ray image as green channel and the red channel was filled with zeros.} 
    \label{data_rgb}
\end{figure}

For training neural networks, we use images without contours and marks. For each candidate cluster, a pair of X-ray and optical PNG images were merged into a single PNG image. As well as our custom network, we use existing architectures, that were designed to take input images with 3 colour channels. In order to achieve this, we grayscale the X-ray and optical images and stack them together as individual channels, leaving one channel empty (zero-filled) to create a single RGB image. Although training of our custom network can be done with any number of input channels, we use the same 3-channel images as the input to the network unless stated otherwise. By default, we construct the input images as follows: the blue channel contains the grayscaled optical image, the green contains the grayscaled X-ray image and the red is filled with a matrix of zeros (\autoref{data_rgb}).

\subsection{Data augmentation}
\label{sec:Data augmentation}
With smaller data sets, the risk of over-fitting increases, resulting in poor generalisation to data outside of the training set. To prevent overfitting, we use data augmentation to reduce the probability that the network will see exactly the same image twice and to essentially increase our training sample size. At each training step, the input image is randomly scaled to a uniform value between $1/1.3$ and $1.3$, rotated by a random uniform angle between 0 and $360^{\circ}$ 
and translated in x and y directions by a random uniform value between $-4$ and $4$ pixels.


\section{Performance measurements}
\label{performance_measurements}

This section describes the measurement methods we chose to evaluate our neural networks compared to a baseline.

Accuracy is the most intuitive performance measurement. It is the ratio of correct predictions to all predictions and is defined as

\begin{equation}
    A = \frac{TP + TN}{TP + TN + FP + FN},
\end{equation}

\noindent where $TP$ refers to the number of true positives, in our case the number of clusters correctly classified as clusters, $TN$ is a number of true negatives (number of non-clusters correctly classified as non-clusters), $FP$ is a number of false positives (number of non-cluster incorrectly classified as clusters) and $FN$ states for a number of false negatives (number of clusters incorrectly classified as non-clusters).

Precision is the ratio of the correctly classified positives (i.e. clusters) and all objects classified as positives. This is defined as

\begin{equation}
    P = \frac{TP}{TP + FP}.
\end{equation}

Recall is the ratio of the correctly classified positives and all positives examples in the test data. It is defined as

\begin{equation}
    R = \frac{TP}{TP + FN}.
\end{equation}

The receiver operating characteristic (ROC) is a performance measurement of detection problems plotted as a true positive rate (recall) against the false positive rate, defined as

\begin{equation}
    FPR = \frac{TN}{TN + FP}
\end{equation}

\noindent at various thresholds. The area under the curve (AUC) describes the model's capability to distinguish between two classification classes and is independent of the choice of the threshold. When reporting detection performance for a class (from the CNN output) in terms of ROC curve, we compare the posterior probability of the class to a varying detection threshold.

\section{Results and Discussion}
\label{sec:Results and Discussion}

\subsection{The Hunt for Galaxy Clusters results}
\label{subsec:The Hunt for Galaxy Clusters results}

The data set of 1\,600 galaxy cluster candidates in \textit{The Hunt for Galaxy Clusters} contained 404 objects previously classified by experts. 

\begin{figure}
	\includegraphics[width=\columnwidth]{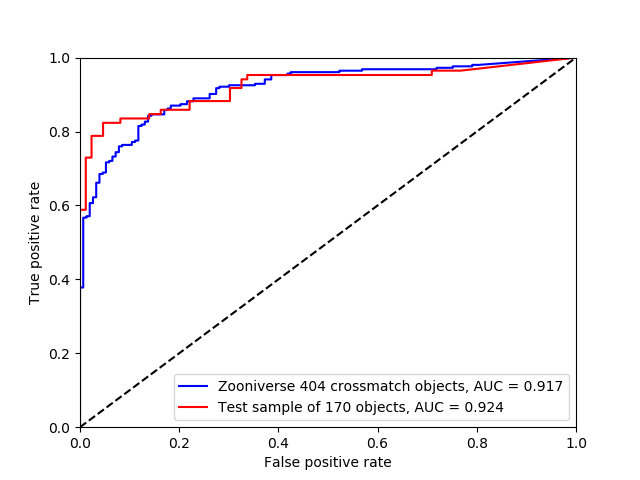}
	\caption{The receiver operating characteristic (ROC) curves for the classifications by Zooniverse volunteers, taking the classifications of experts as the ground truth. Closer the curve copies the left vertical and top horizontal axis, better the classifier. The dashed line shows how would the results be if the people guessed totally randomly.}
	\label{fig:ROC_Zooniverse_concatenated}
\end{figure}

\autoref{table_of_volunteers_classifications_evaluation} displays a comparison of the unweighted and weighted classifications of the Zooniverse volunteers (\autoref{sec_weighting_volunteers_classifications}) based on the agreement with the experts. \autoref{fig:ROC_Zooniverse_concatenated} shows ROC curves computed for the whole crossmatch sample of 404 objects classified by both the Zooniverse volunteers and experts and the ROC computed on a subsample of 170 objects, 85 spectroscopically confirmed galaxy clusters and 85 objects classified as non-clusters by experts. This subsample is also used for the testing of the CNNs. The Zooniverse volunteers performed better on the subsample of 170 objects than on the whole crossmatch sample of 404 objects. This could be an indication of a bias towards correctly classifying easier objects since spectroscopically confirmed galaxy clusters tend to be larger.
\begin{table*}
    \caption{The results of cluster classification by Zooniverse volunteers on two data sets, 404 objects are those classified by both, scientists and Zooniverse volunteers, the 170 objects data set is a subsample of the 404 objects, where 85 objects are spectroscopically confirmed clusters and 85 are objects classified as non-clusters by experts.}

	\begin{center}	
		\begin{tabular}{ll|rrrrrrrr}
			Data set & Zooniverse Classifications & TP & TN & FP & FN & accuracy &  precision & recall  \\
			\hline
			404 objects & unweighted & 69 & 150 & 0 & 185 & 0.542 & 1.000 & 0.272 \\
			404 objects & weighted &  102 & 149 & 1 & 152 & 0.621 & 0.990 & 0.401 \\
			170 objects & weighted & 55 & 84 & 1 & 30 & 0.818 & 0.982 & 0.647 \\
		\end{tabular}	
		\label{table_of_volunteers_classifications_evaluation}
		
	\end{center}
\end{table*}

\begin{figure}
	\includegraphics[width=\columnwidth]{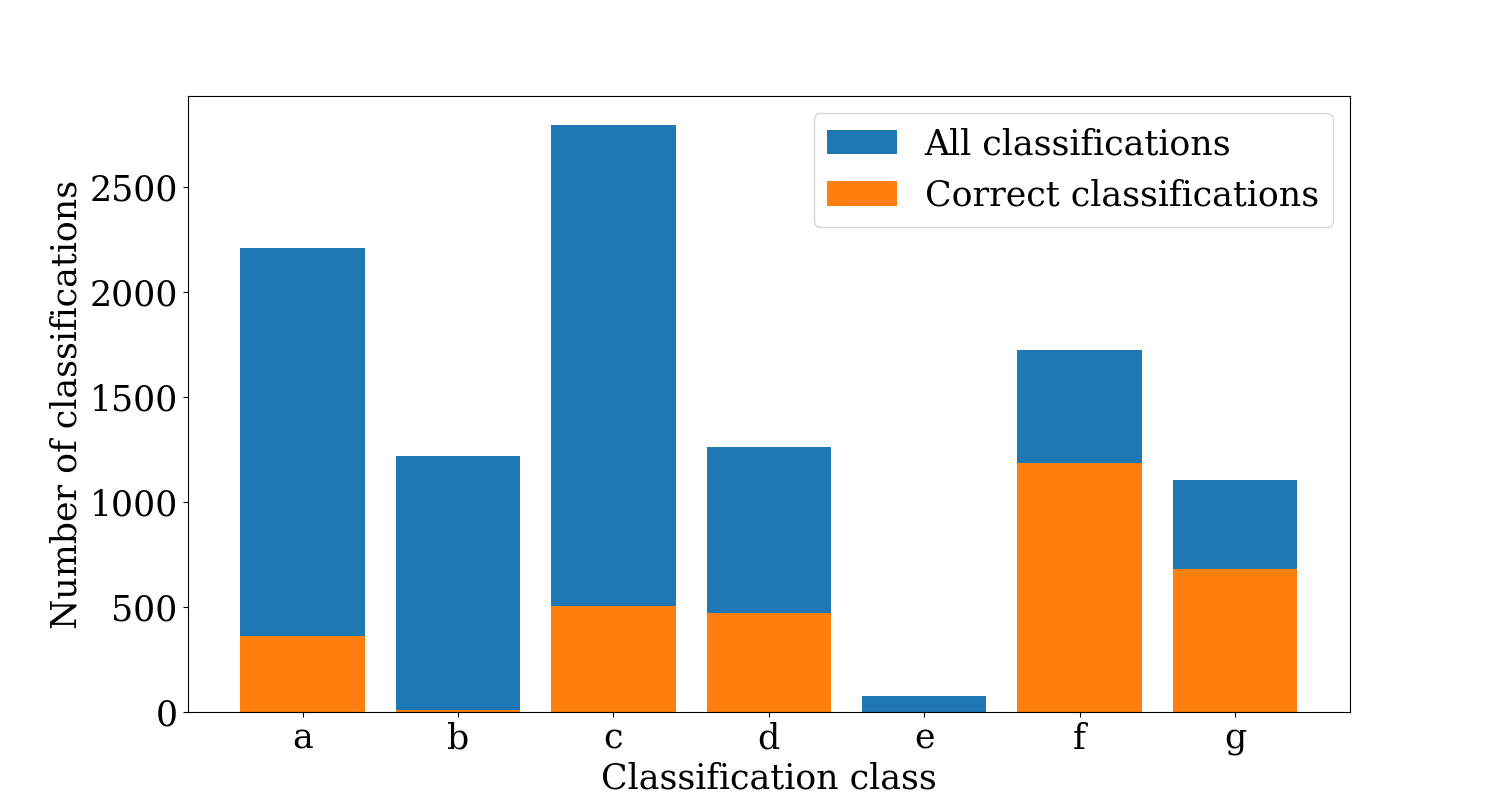}
	\caption{A quantification of the Zooniverse classifications for a)\,no~emission, b)\,edge, c)\,point, d)\,nearby~galaxy, e)\,no~optical~image, f)\,nearby~galaxy~cluster, g)\,distant~galaxy~cluster, assuming the ground truth is the expert classification.}
	\label{fig:histogram_comparison_of_correct_to_incorrect_classifications}
\end{figure}

\autoref{fig:histogram_comparison_of_correct_to_incorrect_classifications} shows the fraction of the Zooniverse volunteer's individual answers in agreement with experts to all Zooniverse answers for classification ending answers, except for \textit{not a nearby cluster} and \textit{not a distant cluster}, which do not have a direct counterpart in the classification of experts. Assuming that the expert classifications are the ground truth, the biggest difficulty for the volunteers seems to be distinguishing extended from point-like X-ray emission. Also, the volunteers inconsistently classified a large fraction of \textit{no emission} classes, suggesting that they struggled to interpret the X-ray images. The huge discrepancy between volunteer's individual classifications and classifications of experts were in the \textit{edge} category, used for galaxy cluster candidates close to the edge of \textit{XMM-Newton's} chips and its field of view. Based upon discussions within the online forum, we assume that this bias could emerge from \textit{XMM-Newton's} grid-like pattern created by small gaps between its individual detectors, which volunteers often mistaken for the edge of the chips. The \textit{nearby galaxy} category was also a difficult question for the volunteers. Again based on the forum discussion we find that volunteers often classified nearby galaxy clusters with a prominent brightest central galaxy as a \textit{nearby galaxy} class, which could lead to many nearby galaxy clusters missed. In general, the Zooniverse volunteers preferentially classified objects as $non-clusters$.

Some of the biases could be mitigated in possible future versions of the project if explanations were clearer and more focus was put on example images in the help notes. Possibly the most important biases were often a classification of an X-ray emission as \textit{no emission} and misclassification of an extended X-ray emission as a point-like X-ray emission. This are the main reasons why clusters were missed by the Zooniverse volunteers.
We tried to keep in mind the possibility of low scientific knowledge of the volunteers and not to overwhelm the volunteers with huge amounts of information, which could discourage them, but we were still able to provide a detailed explanation of the X-ray emission in the tutorial and the help notes, with nice example images and diagrams to help with the X-ray contours.
Small interviews with our beta testers revealed that around 20\,\% of them did not read the supporting texts. It might be possible that classifications with a lot of disagreement in the interpretation of the X-ray emission preferentially came from volunteers who did not adequately read the supporting material. A questionnaire would be needed to further probe this possibility. These biases could be cut down with simpler and shorter explanations of the X-ray properties, so it would be easier to understand and less information to digest.
Another common tendency was the misclassification of nearby clusters that contain prominent BCGs (brightest cluster galaxies), with that of nearby galaxies. This could be reduced with a dedicated pair of images for the two situations in the help notes.

We have to note that even the classifications of experts could be biased towards \textit{low-z} clusters, since we use DSS optical images, which are limited to z\,$\sim$\,0.3.

Another possible bias may come from the fact that spectroscopically confirmed clusters are biased to big clusters, which might affect our interpretation.

To explore if the Zooniverse volunteers were biased finding preferentially most prominent galaxy clusters, we made extent -- extension likelihood plane plots (see \aref{Appendix:Extent -- extent likelihood plots}). We found that the galaxy clusters found by the Zooniverse volunteers populate all of the space, not showing bias and their sample of galaxy clusters also can not be recreated by a simple cut in this space.

Even though the Zooniverse volunteers did not show a high accuracy compared to experts, misclassifying many galaxy clusters as other options, the sample of galaxy clusters they selected is pure. This makes us conclude that, via the Zooniverse project, the general public can help scientific research where a very pure sample of galaxy clusters is required, but it did not prove to be helpful in a case where a sample of galaxy clusters should be complete.

\subsection{CNN training}
\label{CNN training}
We use two different data sets, one classified by experts and one by the Zooniverse volunteers. We use balanced training batches, containing the same number of classification classes, randomly sampled from the training data. This is to prevent the network from being biased towards the class that occurs most frequently in the training sample.

\begin{figure}
	\includegraphics[width=\columnwidth]{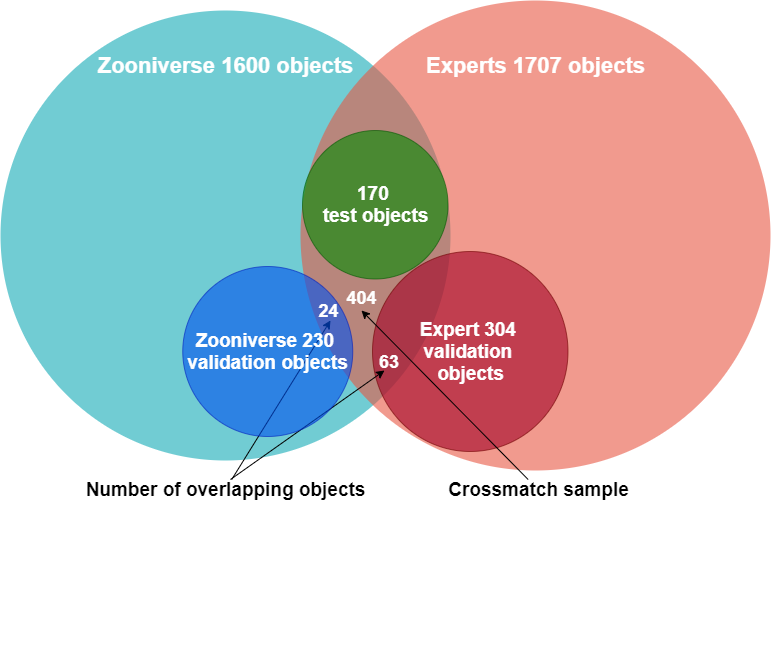}
	\caption{A Venn diagram presenting the data sets.}
	\label{fig:paper_venn_diagram_all_objects}
\end{figure}

Regardless of the training data, all the networks were tested on the same data set of 85 spectroscopically confirmed galaxy clusters and 85 objects classified as non-clusters by the experts, the 170 test objects. \autoref{table_numbers_of_objects_in_train_and_valid} and \autoref{fig:paper_venn_diagram_all_objects} describe the numbers of objects used in the training, validation and test data sets, classified by experts and the Zooniverse volunteers for testing on the 170 object test sample. All the networks were trained on grayscaled and combined X-ray and optical images as described in \autoref{Data preprocessing} if not stated otherwise.

\begin{table}
	\caption{The number of objects in the training, validation and test data sets classified by Zooniverse and experts.}
	\begin{center}
		\begin{tabular}{l|rr|rrr} 
			\multirow{2}{*}{Class} & \multicolumn{2}{c}{Zooniverse} & \multicolumn{2}{c}{Experts} \\ \cline{2-6}
			 & Train & Validate & Train & Validate & Test\\
			 \hline
			 cluster & 320 & 130 & 845 & 200 & 85\\
			 non-cluster & 880 & 100 & 388 & 104 & 85\\
			 total & 1200 & 230 & 1233 & 304 & 170
		\end{tabular}
	\end{center}
	\label{table_numbers_of_objects_in_train_and_valid}
\end{table}

\begin{table}
	\caption{The architecture of our custom network which achieved the best performance. Each of the convolutional and dense layers is followed by a ReLU non-linearity with the exception of the final output dense layer which has the softmax for classification.}

    \centering
    \begin{tabular}{c|lccc}
          Layer & Layer type & filter shape / stride & input shape \\
         \hline
         1 & conv &  3$\times$3$\times$64/(1, 1) & 356$\times$356$\times$3 \\
         2 & max pool & 2$\times$2/(2, 2) & 356$\times$356$\times$64 \\
         3 & conv & 3$\times$3$\times$32/(1, 1) & 178$\times$178$\times$64 \\
         4 & max pool & 2$\times$2/(2, 2) & 178$\times$178$\times$32 \\
         5 & conv & 3$\times$3$\times$32/(1, 1) & 89$\times$89$\times$32 \\
         6 & max pool & 2$\times$2/(2, 2) & 89$\times$89$\times$32 \\
         7 & conv & 3$\times$3$\times$32/(1, 1) & 45$\times$45$\times$32 \\
         8 & max pool & 2$\times$2/(2, 2) & 45$\times$45$\times$32 \\
         9 & conv & 3$\times$3$\times$32/(1, 1) & 23$\times$23$\times$32 \\
         10 & max pool & 2$\times$2/(2, 2) & 23$\times$23$\times$32 \\
         11 & conv & 3$\times$3$\times$32/(1, 1) & 12$\times$12$\times$32 \\
         12 & max pool & 2$\times$2/(2, 2) & 12$\times$12$\times$32 \\
         13 & flatten & - & 6$\times$6$\times$32  \\
         14 & dense & 256 & 1152  \\
         15 & dense & 2 & 256 \\
         
    \end{tabular}
    \label{tab:custom_net_architecture}
\end{table}

We experimented with both a custom network (\autoref{tab:custom_net_architecture}) and using 3 different state of the art CNN architectures: VGG19~\citep{Simonyan2014}, InceptionV3~\citep{Szegedy2015} and MobileNet~\citep{Howard2017}. We used those networks with their pre-trained weights, using a large learning rate and unfreezing all the layers. Of the 3 models, MobileNet, pre-trained on the ImageNet \citep{Deng2009}, achieved the best performance and therefore we only discuss this architecture. Similarly, \citet{Lieu2018} found MobileNet to be the superior architecture for classifying solar system objects. The hyperparameters for our custom network and the MobileNet network are given in \autoref{table_of_hyperparameters}. We used Keras \citep{chollet2015} with TensorFlow \citep{tensorflow2015} backend. The \texttt{lr. red. patience} and \texttt{lr. red. factor} are parameters of the \texttt{ReduceLROnPlateau} Keras callback. The parameter \texttt{lr. red. patience} defines how many epochs without improvement of the validation accuracy (different proxy can be chosen to monitor) have to pass to change the current learning rate by multiplying it with the \texttt{lr. red. factor}.

The batches used to train the networks were randomly generated during training, always from the whole training sample. Validation started once a satisfying number of generated batches was presented to the network, this is the training data set size divided by the batch size. This was done to maximise the use of our data while keeping balanced numbers of classes in the yielded training batches, in order to avoid biasing the network.

\begin{table}
    \caption{Hyperparameters of our custom network and the MobileNet network. The number of iterations, batches yielded during training, is shown for training on the data set classified by experts.}
	\begin{center}	
		\begin{tabular}{c|ccc}		
		    Hyperparameters & Custom net & MobileNet \\
			\hline
			Batch size & 10 & 20 \\
		    Iterations & 153\,000 & 3\,825 \\
		    Optimizer & SGD & Adadelta \\
		    Nest. Momentum & 0.90 & - \\
		    Rho & - & 0.95 \\
		    Initial lr. & 0.0001 & 1.0 \\
		    lr. decay & 10$^{-6}$ & 0.95 \\
		    Minimal lr. & 10$^{-4}$ & 0.01 \\
		    lr. red. patience & 14 & 4 \\
		    lr. red. factor & 0.75 & 0.85 \\
		    Dense dropout & 0.65 & 0.65 \\
		    Output activation & softmax & softmax \\
		    Loss function & cat. crossentropy & cat. crossentropy \\
		    Input image size & 356$\times$356 & 224$\times$224 \\

		\end{tabular}	
		\label{table_of_hyperparameters}
	\end{center}
\end{table}

\subsection{CNN results}
\label{CNN results}
\begin{figure}
	\includegraphics[width=\columnwidth]{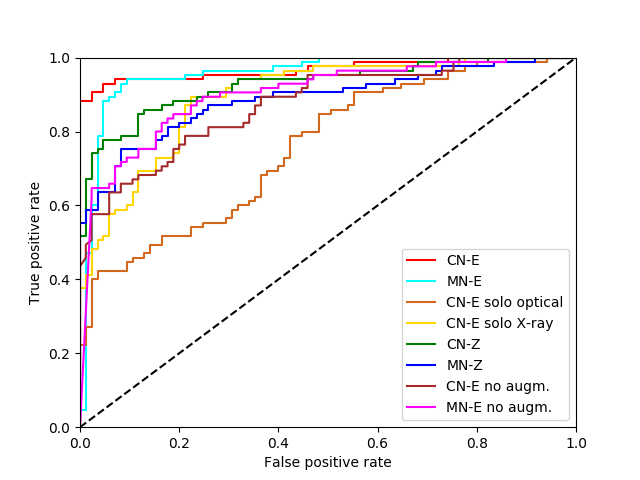}
	\caption{ROC curves for the best-performing networks when trained on different data formats. Closer the curve copies the left vertical and top horizontal axis, better the classifier. The dashed line represents how would an untrained, randomly guessing classifier score. Training on optical data only ended up with the poorest results, using only X-ray data achieved much better results, however, the combination of optical and X-ray data resulted in the best performance. CN refers to our custom network, MN to the MobileNet architecture, E to the data set classified by experts, Z to the data set classified by the Zooniverse volunteers.}
	\label{fig:concatenated_ROC_in_one_figure_6_networks}
\end{figure}

We demonstrate that convolutional neural networks are capable of high accuracy, automated galaxy cluster candidate classification. We trained each of our networks 10 times with the exact same hyperparameters, differing only in the seed for generation of random numbers during network's initialisation, the order of random image selection into balanced mini-batches during training and the random sampling of augmentation values applied during training but keeping the same objects in the training, validation and test data sets. The results of individual runs are averaged and presented together with their standard deviations in \autoref{tab:CNN_all_averaged_results_std} and \autoref{fig:concatenated_ROC_in_one_figure_6_networks}, helping to compare various networks.

To report accuracy (A), precision (P) and recall (R) in \autoref{tab:CNN_all_averaged_results_std}), we need to make hard classification decision for each example image from our test set. Our neural networks are trained to output the probability that the input image is a galaxy cluster. Therefore, we classify input images as galaxy cluster if this probability is higher than 0.5.

\begin{table*}
\caption{Averaged galaxy cluster candidate classification results of the networks each trained 10 times with the exact same hyperparameters, only with a different seed for generation of random numbers during its initialisation.}
\begin{center}
\begin{tabular}{c | cccc } 

network & A$\,\pm\,$std & P$\,\pm\,$std & R$\,\pm\,$std & AUC\,$\pm$\,std \\
\hline
CN-E & 0.90\,$\pm$\,0.03 & 0.89\,$\pm$\,0.05 & 0.91\,$\pm$\,0.03 & 0.96\,$\pm$\,0.01 \\
MN-E & 0.88\,$\pm$\,0.02 & 0.87\,$\pm$\,0.03 & 0.91\,$\pm$\,0.03 & 0.94\,$\pm$\,0.01 \\
CN-E solo optical & 0.68\,$\pm$\,0.02 & 0.64\,$\pm$\,0.02 & 0.85\,$\pm$\,0.04 & 0.77\,$\pm$\,0.02 \\
CN-E solo x-ray & 0.81\,$\pm$\,0.01 & 0.78\,$\pm$\,0.03 & 0.86\,$\pm$\,0.04 & 0.89\,$\pm$\,0.01 \\
CN-Z & 0.82\,$\pm$\,0.01 & 0.96\,$\pm$\,0.01 & 0.67\,$\pm$\,0.02 & 0.91\,$\pm$\,0.01 \\
MN-Z & 0.79\,$\pm$\,0.02 & 0.96\,$\pm$\,0.03 & 0.62\,$\pm$\,0.03 & 0.86\,$\pm$\,0.02 \\
CN-E no augm. & 0.75\,$\pm$\,0.02 & 0.70\,$\pm$\,0.02 & 0.87\,$\pm$\,0.03 & 0.87\,$\pm$\,0.01 \\
MN-E no augm. & 0.81\,$\pm$\,0.01 & 0.75\,$\pm$\,0.02 & 0.91\,$\pm$\,0.01 & 0.90\,$\pm$\,0.02 \\

\end{tabular}
\label{tab:CNN_all_averaged_results_std}
\end{center}
\end{table*}

Our best-performing custom network (CN-E), trained on the expert classified data set, achieved an average accuracy of (90\,$\pm$\,3)\,\%. We also explored training on concatenated PNG images, without the grayscaling, so having six channels instead of three, but this did not change the performance significantly. 

The MobileNet architecture trained on the data classified by experts achieved an average accuracy of (88\,$\pm$\,2)\,\%. Perhaps MobileNet has slightly different sensitivity for individual colour channels due to the potential bias in its original training sample. We explored this possibility by training it on two additional channel configurations, X-ray green, optical red, empty blue and X-ray red, optical green, empty blue, but its performance did not change significantly.

Training using the labels obtained in the Zooniverse project resulted in lower performance for both, our custom network (CN-Z) and the MobileNet (MN-Z), achieving average accuracies (82\,$\pm$\,1)\,\% and (79\,$\pm$\,2)\,\%, respectively.

Lastly, we also explored the training of neural networks on single wavelength PNG images. Our custom network using expert labels trained only on the X-ray images without their optical counterparts (CN-E solo X-ray) achieved an average accuracy of (81\,$\pm$\,1)\,\%. Our custom network using expert labels trained only on the optical images (CN-E solo optical) performed the worse, achieving an accuracy of only 68\,$\pm$\,2)\,\%. 
This is rather easily understandable knowing that the \textit{XMM-Newton} data are much deeper than the POSS-II images used for the current analysis: while \textit{XMM-Newton} can detect galaxy clusters as extended sources out to z = 1 at least, the POSS sensitivity strongly drops beyond z\,$\sim$\,0.3 rendering galaxies are hardly identifiable.

Using augmentation (\autoref{sec:Data augmentation}) was critical to achieving good performance, the accuracy of the network CN grayscale would drop from (90\,$\pm$\,3)\,\% to (75\,$\pm$\,2)\,\% without the augmentation and from (88\,$\pm$\,2)\,\% to (81\,$\pm$\,1)\,\% for MobileNet.

\subsection{Interpreting the results}
\label{Interpreting the results}

\begin{figure}
	\centering
	
	
	
	
	
	\includegraphics[width=0.15\textwidth]{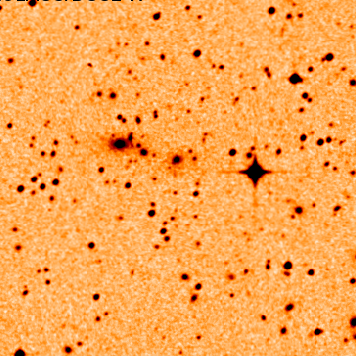}
	\hfill
	\includegraphics[width=0.15\textwidth]{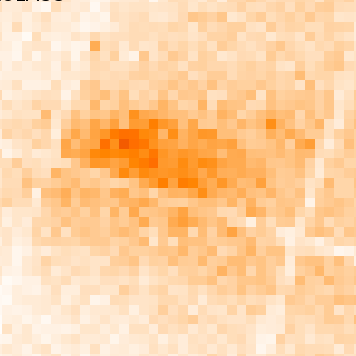}
	\hfill
	\includegraphics[width=0.15\textwidth]{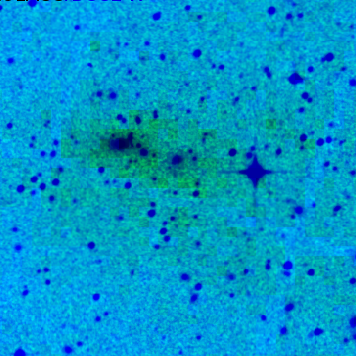}
	
	\vspace{0.01\textwidth}
	
	\includegraphics[width=0.15\textwidth]{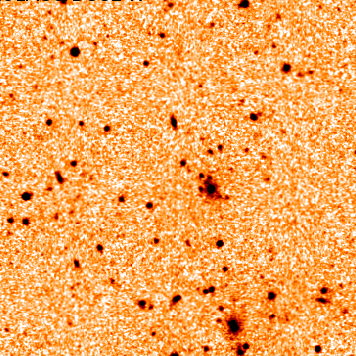}
	\hfill
	\includegraphics[width=0.15\textwidth]{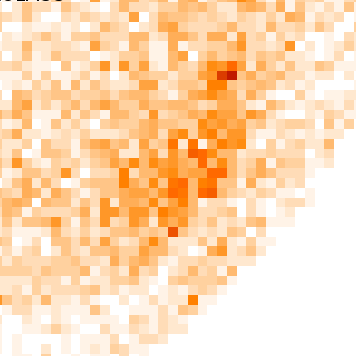}
	\hfill
	\includegraphics[width=0.15\textwidth]{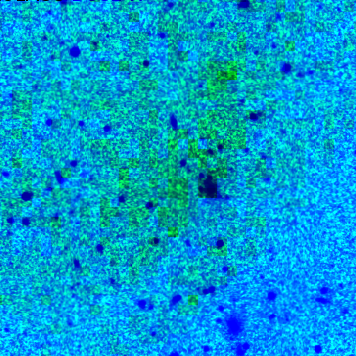}
	
	
	
	
	
	
%
	\vspace{0.01\textwidth}
	
	\includegraphics[width=0.15\textwidth]{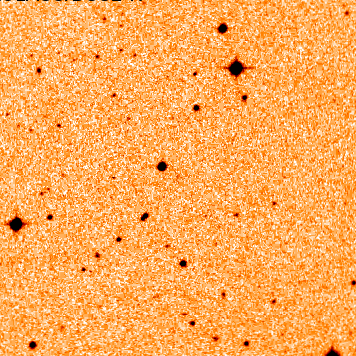}
	\hfill
	\includegraphics[width=0.15\textwidth]{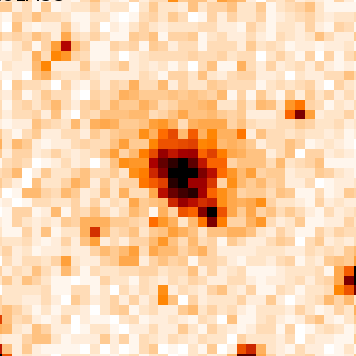}
	\hfill
	\includegraphics[width=0.15\textwidth]{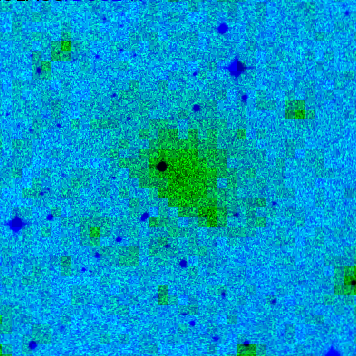}

	\caption{Spectroscopically confirmed galaxy clusters correctly classified by our custom network randomly selected from the test sample (TP). \textit{Left:} optical, \textit{middle:} X-ray, \textit{right:} combined. }
	\label{TP_galaxy_clusters}
\end{figure}

We further investigate the results of the best training run of our custom network (CN-E), which can classify even faint clusters and those close to the edge of \textit{XMM-Newton}'s field of view. \autoref{TP_galaxy_clusters} shows some of these randomly selected correctly classified galaxy clusters.

\begin{figure}
	\centering
	\includegraphics[width=0.15\textwidth]{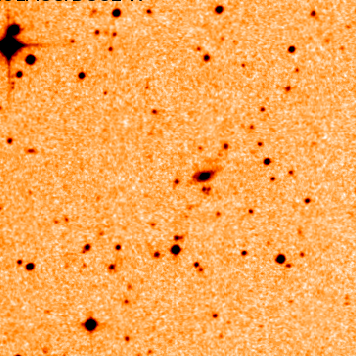}
	\hfill
	\includegraphics[width=0.15\textwidth]{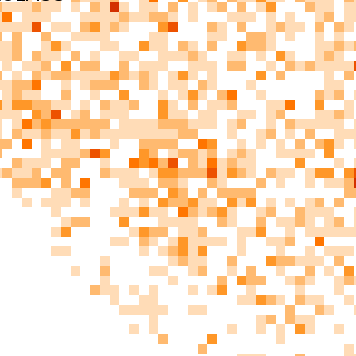}
	\hfill
	\includegraphics[width=0.15\textwidth]{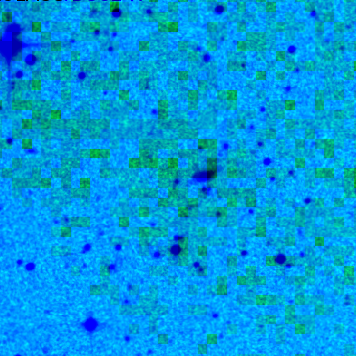}
	
	\vspace{0.01\textwidth}
	
	\includegraphics[width=0.15\textwidth]{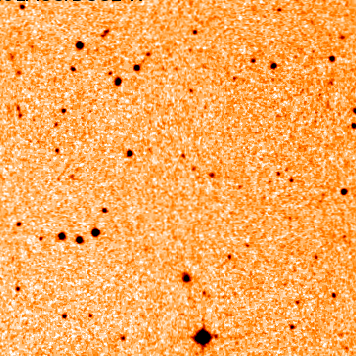}
	\hfill
	\includegraphics[width=0.15\textwidth]{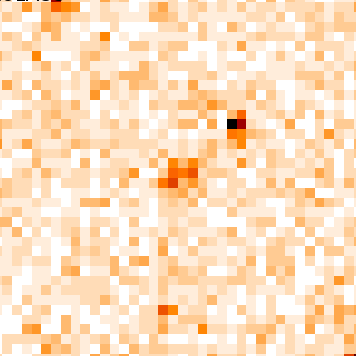}
	\hfill
	\includegraphics[width=0.15\textwidth]{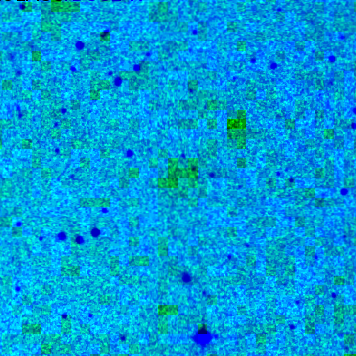}

	\caption{Non-galaxy clusters incorrectly classified as galaxy clusters (FP) by our custom network. \textit{Left:} optical, \textit{middle:} X-ray, \textit{right:} combined.}
	\label{custom_net_FP}
\end{figure}

\begin{figure}
	\centering
	\includegraphics[width=0.14\textwidth]{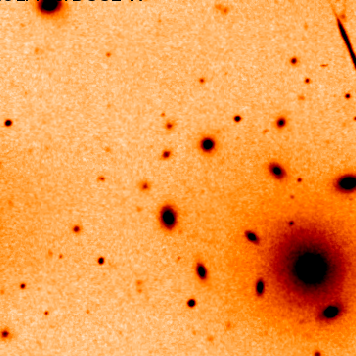}
	\hfill
	\includegraphics[width=0.14\textwidth]{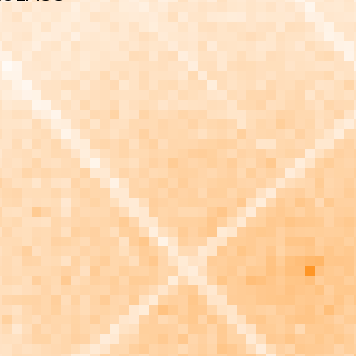}
	\hfill
	\includegraphics[width=0.14\textwidth]{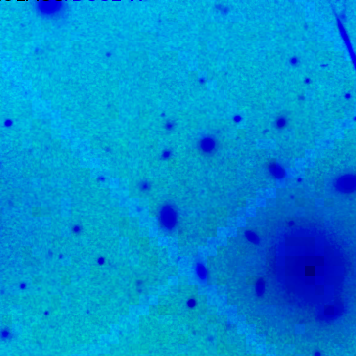}
	
	\vspace{0.01\textwidth}
	
	\includegraphics[width=0.14\textwidth]{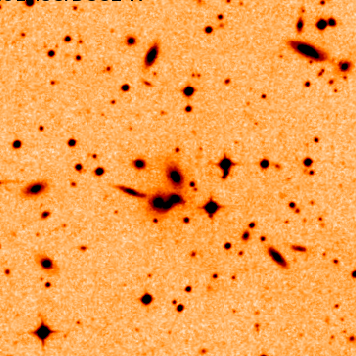}
	\hfill
	\includegraphics[width=0.14\textwidth]{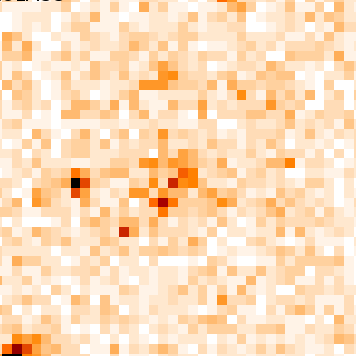}
	\hfill
	\includegraphics[width=0.14\textwidth]{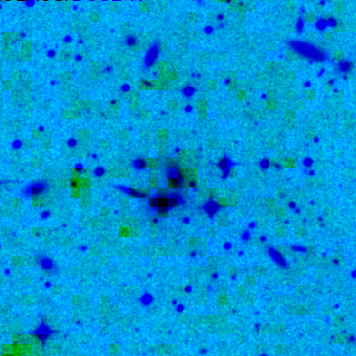}
	
	\vspace{0.01\textwidth}
	
	
	
	
	
	\includegraphics[width=0.14\textwidth]{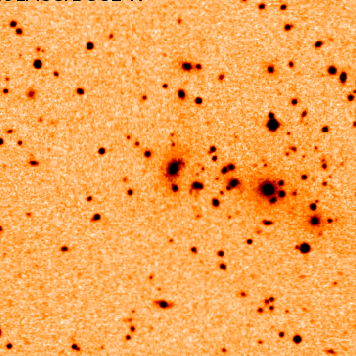}
	\hfill
	\includegraphics[width=0.14\textwidth]{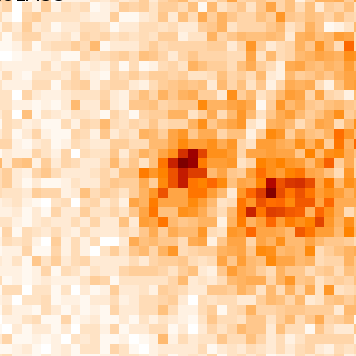}
	\hfill
	\includegraphics[width=0.14\textwidth]{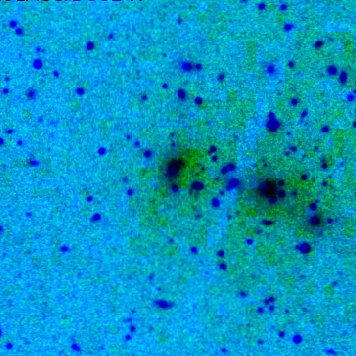}
	
	\vspace{0.01\textwidth}
	
	
	
	\includegraphics[width=0.14\textwidth]{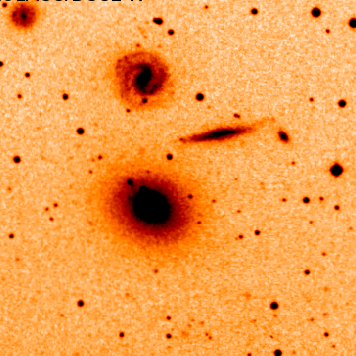} 
	\hfill
	\includegraphics[width=0.14\textwidth]{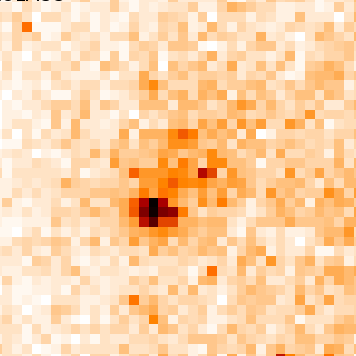}
	\hfill
	\includegraphics[width=0.14\textwidth]{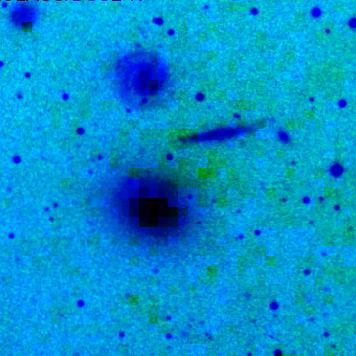}
	
	\vspace{0.01\textwidth}
	
	\includegraphics[width=0.14\textwidth]{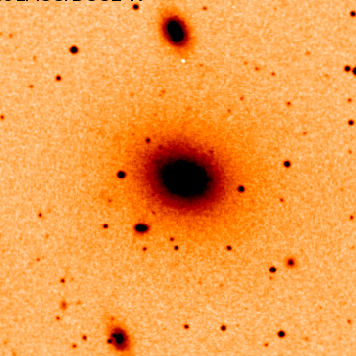}
	\hfill
	\includegraphics[width=0.14\textwidth]{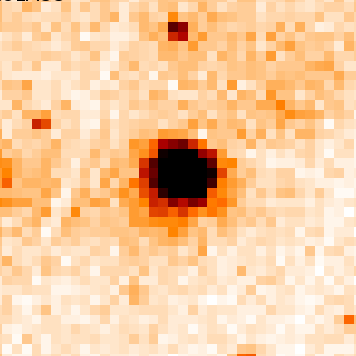}
	\hfill
	\includegraphics[width=0.14\textwidth]{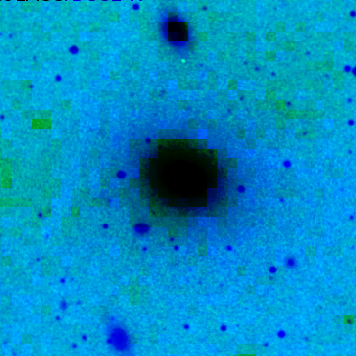}
	
	\caption{Galaxy clusters incorrectly classified as non-galaxy clusters (FN) by our custom network. \textit{Left:} optical, \textit{middle:} X-ray, \textit{right:} combined.}
	\label{custom_net_FN}
\end{figure}

\autoref{custom_net_FP} shows two objects classified as non-clusters by the experts, but as clusters by our custom network. The top object raised a concern that it was actually a galaxy cluster. We assume that it was classified as a galaxy cluster by our custom network because of the presence of the faint X-ray emission in the centre and that it is a promising candidate for further investigation and spectroscopic redshift confirmation. 
\autoref{custom_net_FN} displays images of spectroscopically confirmed galaxy clusters which have been incorrectly classified by our custom network as a \textit{non-cluster} class.
The first object from the top is a non-centered galaxy cluster. The second contains a group of nearby galaxies with faint extended X-ray emission, which might have fooled our network. The third is a cluster that falls on a chip gap. The fourth is a galaxy cluster with three prominent nearby galaxies along the line of sight which is probably what fooled our network, and the last object appears like a nearby galaxy, which can be hard to classify even for the experts.

\begin{figure}
    \centering
    \includegraphics[width=0.145\textwidth]{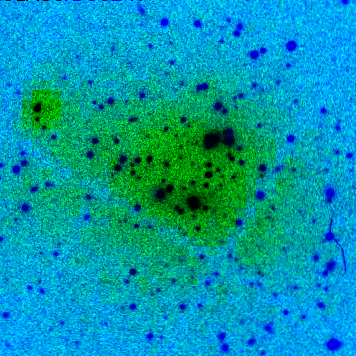}
    \vspace{0.01\textwidth}
    
    
    \includegraphics[width=0.155\textwidth]{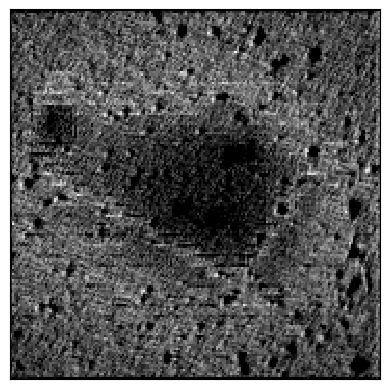}
    \hfill
    \includegraphics[width=0.155\textwidth]{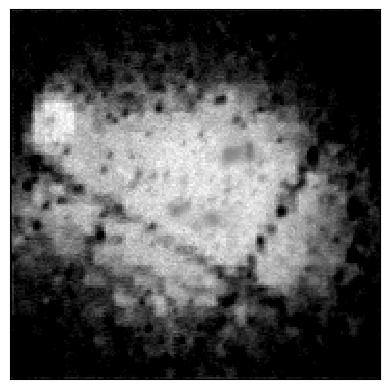}
    \hfill
    \includegraphics[width=0.155\textwidth]{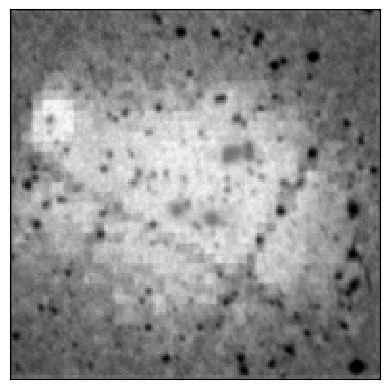}
    
    
    \includegraphics[width=0.155\textwidth]{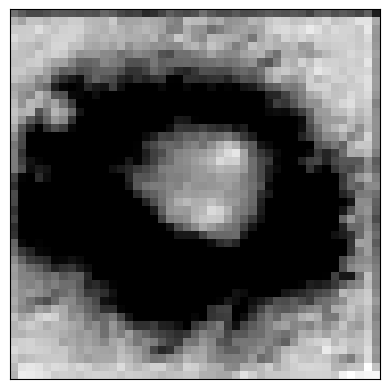}
    \hfill
    \includegraphics[width=0.155\textwidth]{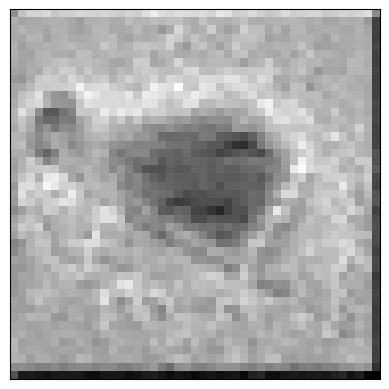}
    \hfill
    \includegraphics[width=0.155\textwidth]{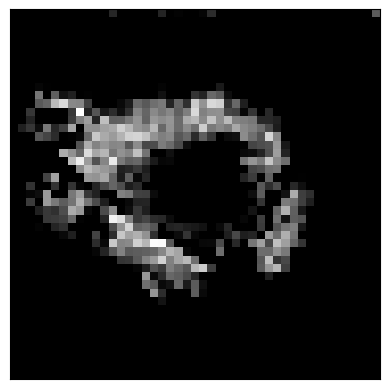}
    
    
    \includegraphics[width=0.155\textwidth]{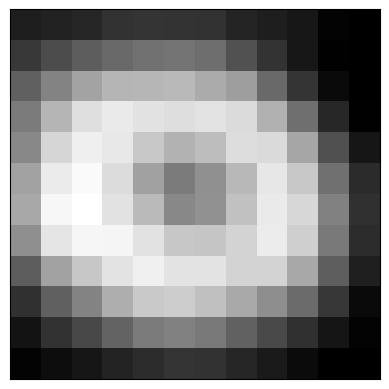}
    \hfill
    \includegraphics[width=0.155\textwidth]{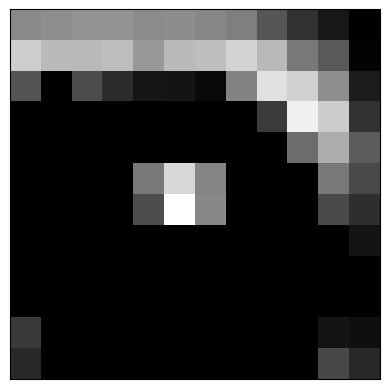}
    \hfill
    \includegraphics[width=0.155\textwidth]{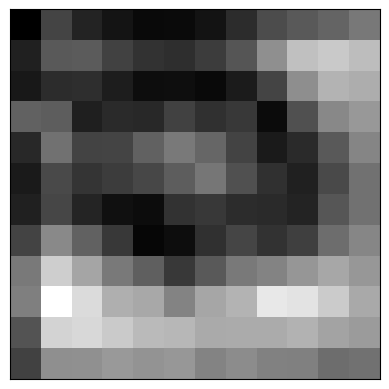}
    
    \caption{Top: Input image to the trained network. Each row from second to last shows outputs (activation maps) of 3 selected filters from 2nd, 4th and 6th convolutional layer of our custom network, respectively.}
    \label{custom_net_activations}
\end{figure}

\autoref{custom_net_activations} shows outputs of the selected filters of our custom network for a spectroscopically confirmed nearby galaxy cluster. We can see how the network learned to search for edges and colour patches of X-ray or optical light. Some filters learned to search primarily for X-ray emission and others for optical emission. Most of the filters detected both of the emission components simultaneously. Multiple filters in the same layer usually learned to search for X-ray emission, but their sensitivity is different. There are filters which get activated only by stronger emission, while other filters are more sensitive to X-ray emission. The network uses the filters to probe the presence and extent of the X-ray emission in the input image. Note that the filter output size decreases deeper within the network because of the max-pooling operation applied in the pooling layer after each convolutional layer.

\begin{table}
	\caption{Results from the multi-class classification networks.}
	\begin{center}	
		\begin{tabular}{c|ccccc}
			class & $A$ & $P$ & $R$ & AUC\\
			\hline
			MN grayscale \\ \cline{1-1} 
			Low-z cluster & 0.77 & 0.62 & 0.94 & 0.93 \\
			High-z cluster & 0.87 & 0.56 & 0.22 & 0.91 \\
			Point source & 0.87 & 0.88 & 0.36 & 0.89 \\
			Nearby galaxy & 0.90 & 0.70 & 0.73 & 0.92 \\
			Other & 0.91 & 0.65 & 0.68 & 0.92 \\
			\hline
			CN grayscale \\ \cline{1-1}
			Low-z cluster & 0.79 & 0.68 & 0.81 & 0.89 \\
			High-z cluster & 0.84 & 0.44 & 0.65 & 0.89 \\
			Point source & 0.84 & 0.75 & 0.27 & 0.88 \\
			Nearby galaxy & 0.89 & 0.74 & 0.57 & 0.85 \\
			Other & 0.87 & 0.52 & 0.64 & 0.88 \\
		\end{tabular}	
	\end{center}
	\label{results_multiclass}
\end{table}

\subsection{Multi-class classification}
\label{Multi-class classification}

We also trained neural networks for multi-class classification using the labels of the experts. We  segregated objects into 5 classification classes - $low~z~cluster$, $high~z~cluster$, $nearby~galaxy$, $point~source$ (point, star or AGN, double source) and $other$ (artefact, edge). The ROC curves and performance measurements were calculated as one versus all problem.

In this regime, the MobileNet architecture and our custom network achieved an AUC and accuracy, averaged over all classes, within 1 sigma. The MobileNet achieved an AUC score of (91\,$\pm$\,2)\,\% and accuracy of (86\,$\pm$\,6)\,\% , and our custom network obtained an AUC of (88\,$\pm$\,2)\,\% and (85\,$\pm$\,4)\,\% accuracy (\autoref{results_multiclass}).

\begin{figure*}
\centering
	\includegraphics[width=\columnwidth]{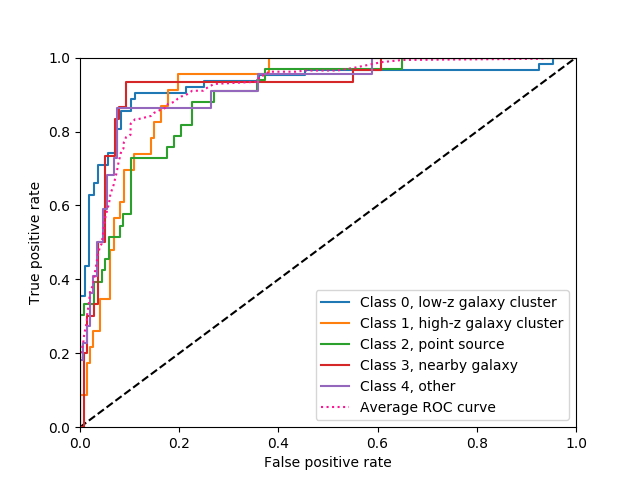}
	\includegraphics[width=\columnwidth]{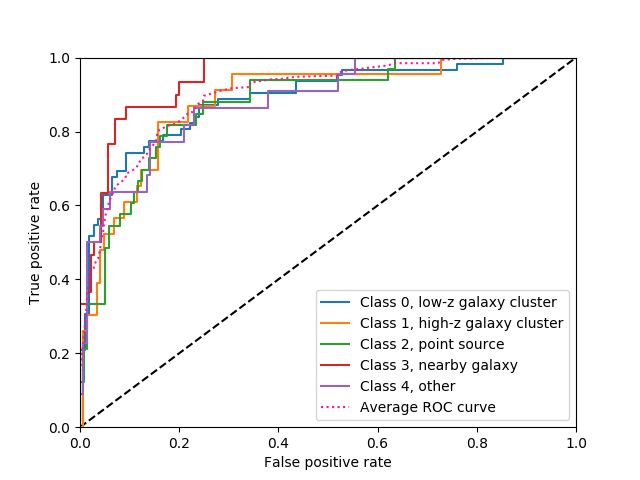}
	\caption{ROC curves for multi-class classification performed by the MobileNet architecture (\textit{left}) and our custom network (\textit{right}).}
	\label{fig:ROC_multiclass}
\end{figure*}

In the case of multi-class classification problems, ROC and AUC are plotted for each of the classes separately as one versus all, reducing the problem to the binary case. From the ROC curves (\autoref{fig:ROC_multiclass}), we see that the \textit{point source} and \textit{high-z galaxy cluster} were the hardest classes to detect, and in the custom network, the \textit{nearby galaxy} class was the easiest to distinguish. We interpret this as a consequence of nearby galaxies being very distinct from the other classes in the optical. Interestingly, this category did not achieve the best performance for the MobileNet network, however, it still placed among the top-performing classes.

We note that since we have trained the neural networks on a sample of galaxy cluster candidates picked by the \xamin pipeline, our sample of point sources is biased towards objects with some spatially extended emission. Thus we can not consider the networks trained for multi-class classification as a reliable point source classifiers since they are not representative of the population and do not reflect the typical appearance of an X-ray point source. 
If one would like to use our neural networks for 
point source detection, re-training or fine-tuning of our models on a representative sample of X-ray point sources would be required.

\subsection{Cross-validation}
\label{sec:Cross-validation}

\begin{table}
	\caption{The number of objects in the training, validation and test data sets in a single fold of the 10 fold cross-validation.}
	\begin{center}
		\begin{tabular}{l|rrr} 
			\multirow{2}{*}{Class} & \multicolumn{3}{c}{Experts} \\ \cline{2-4}
			 & Train & Validate & Test\\
			 \hline
			 cluster & 904 & 113 & 113 \\
			 non-cluster & 399 & 57 & 114 \\
		\end{tabular}
	\end{center}
	\label{table_numbers_of_objects_in_single_fold_crossvalidation}
\end{table}

\begin{figure}
	\includegraphics[width=\columnwidth]{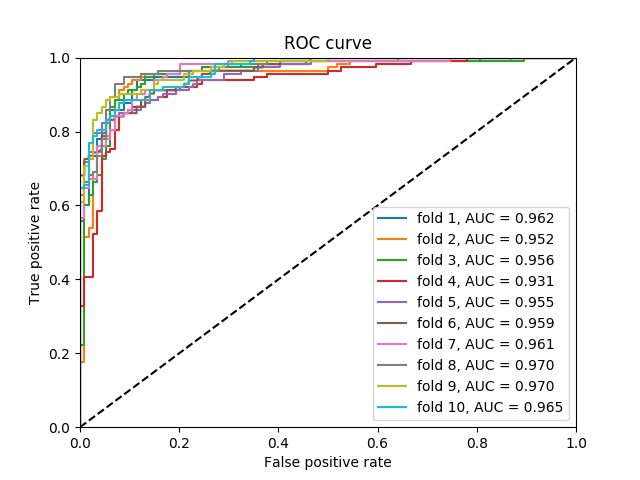}
	\caption{ROC curves for 10 fold cross-validation of our custom network trained on expert classifications.}
	\label{fig:ROC_crossvalidation}
\end{figure}

\begin{table}
\caption{Classification results of our custom networks for a 10 fold cross-validation on classifications done by experts.}
\begin{center}
\begin{tabular}{c | ccc }
Fold & A & P & R \\
\hline
1 & 0.89 & 0.89 & 0.88 \\
2 & 0.92 & 0.91 & 0.93 \\
3 & 0.90 & 0.90 & 0.91 \\
4 & 0.88 & 0.91 & 0.83 \\
5 & 0.87 & 0.88 & 0.86 \\
6 & 0.87 & 0.87 & 0.88 \\
7 & 0.88 & 0.90 & 0.86 \\
8 & 0.92 & 0.89 & 0.95 \\
9 & 0.88 & 0.84 & 0.94 \\
10 & 0.89 & 0.92 & 0.87 \\

\end{tabular}
\label{tab:Crossvalidation}
\end{center}
\end{table}

We perform 10-fold cross-validation of CN-E to explore, if the test data set, having all of its galaxy clusters spectroscopically confirmed, shows significant bias compared to the galaxy cluster sample in the training data set.
\autoref{table_numbers_of_objects_in_single_fold_crossvalidation} contains the number of example images in each data set for a single fold of the cross-validation.
The cross-validation accuracy scores between 87\,\% and 92\,\% (\autoref{tab:Crossvalidation}, \autoref{fig:ROC_crossvalidation}) and our CN-E achieved accuracy 90\,\% on average (\autoref{tab:CNN_all_averaged_results_std}, \autoref{fig:concatenated_ROC_in_one_figure_6_networks}). Those results are consistent and the test sample we used does not seem to have any significant bias on the network's performance.

\section{Summary}
\label{sec:Conclusions}
In this paper, we have presented convolutional neural networks to classify extended X-ray sources detected by the \xamin pipeline. This automated method can be used to replace the traditional manual screening confirmation task of the \xamin galaxy cluster candidates, which is often tedious and slow. 

Firstly, we built a crowd-sourcing Zooniverse project~-~\textit{The Hunt for Galaxy Clusters}, to obtain a classification of a large number (1\,600) of galaxy cluster candidates in a short time frame (6 months). Our volunteers obtained 62$\%$ agreement with experts for identifying clusters and non-clusters in an overlapping sample of 404 objects. We found that the volunteers were often incorrectly classifying objects as point sources or no emission. 
Out of 254 objects classified as galaxy clusters by experts in the overlapping sample, volunteers agreed on 104 of those (66/146 \textit{low-z} and 38/108 \textit{high-z} galaxy clusters), which is only about 40\,\%, but they inconsistently classified only 1 non-cluster as a galaxy cluster. In total, the volunteers found 506 clusters from 1\,600 candidates.
We suspect the reason behind this low performance of the Zooniverse volunteers in \textit{The Hunt for Galaxy Cluster}, if compared to e.g. Galaxy Zoo, to be the complexity of combined X-ray and optical data of galaxy cluster candidates, burdened by multiple projection and instrumental effects (see \autoref{subsec:The Hunt for Galaxy Clusters results} for discussion of biases the Zooniverse volunteers exhibited). We also tested a hypothesis, that the Zooniverse volunteers would preferentially find prominent galaxy clusters and that their sample could be easily recreated by a cut in the extent -- extension likelihood plane \citep{Pacaud2006}, however, the Zooniverse volunteers found galaxy clusters across the entire extent -- extension likelihood space (\aref{Appendix:Extent -- extent likelihood plots}), pointing out that their help could be used for a galaxy cluster science.

Next, we trained CNNs on \textit{XMM-Newton} X-ray images combined with their optical counterparts from DSS2, to distinguish galaxy clusters from non-clusters. 
The cross-validation of our custom network shows consistent results (\autoref{tab:Crossvalidation}, \autoref{fig:ROC_crossvalidation}) with accuracy scoring between 87\,\% and 92\,\%. We further developed networks on a fixed training, validation and test samples, the networks trained on Zooniverse classified data having a different training and validation samples than those trained on data classified by experts, but both having the same test sample.
Our best network (CN-E) obtained an average accuracy of 90\,$\%$ (\autoref{CNN results}). This network used our custom architecture and was trained on labels made by experts. The test sample of 170 objects is composed of 85 spectroscopically confirmed galaxy clusters (62 \textit{low-z} and 23 \textit{high-z}), and 85 galaxy cluster candidates classified as non-clusters by experts. For comparison, a similar network using the MobileNet architecture (MN-E) obtained an average accuracy of 88\,$\%$ and using the custom architecture with the Zooniverse classifications (CN-Z) gave an average accuracy of 82\,$\%$ at best.

In this work, we show that CNNs trained using either X-ray only or optical only images had significantly lower performance in reliably identifying galaxy clusters in comparison to using the combined data. While in the X-rays \textit{XMM-Newton} detects galaxy clusters as extended sources to z = 1 at least, the optical POSS-II data sensitivity strongly drops beyond z\,$\sim$\,0.3, making galaxies hardly identifiable. This is evident from the high number of false-positive detections of galaxy clusters (low precision) using the optical only data. The X-ray only network achieved higher accuracy (81\,\%) than the optical only network (68\,\%).

Additionally we train our networks for multi-class classification using expert classified labels: \textit{low-z galaxy cluster}, \textit{high-z galaxy cluster}, \textit{point source}, \textit{nearby galaxy} and \textit{other}. In this case, the MobileNet architecture performed slightly, but not significantly, better than our custom network (\autoref{tab:CNN_all_averaged_results_std}).

This project is a pilot study to determine the potential of CNNs for the detection of galaxy clusters. In the future, we intend to apply our methods to large sky surveys such as the new \textit{eROSITA} or LSST and \textit{Euclid}. Their enormous data sets are expected to contain tens of thousands of new galaxy clusters, which will require automated, fast and reliable methods to identify, as human screening of such large data volumes will be impossible. Our methods can also be applied to simulated data. Our custom network can be easily fine-tuned to, e.g., \textit{eROSITA} simulations and deliver an automated search tool for galaxy clusters from X-ray images. Applying our CNN on simulations will also enable modelling of the cluster selection function, important for cosmological studies, which cannot be done with clusters selected by human inspection due to their inconsistent biases.

\section*{Acknowledgements}

We would like to thank the referee Dr Florence Durret for the valuable comments that helped to improve the paper.

We would like to acknowledge the scientists from the X-CLASS collaboration who manually classified the \xamin\ galaxy cluster candidates, mainly Jean-Baptiste Melin, one of the moderators overseeing all classifications and Edoardo Cucchetti. We are also very thankful to all of our citizen volunteers who participated in \textit{The Hunt for Galaxy Clusters}. We use data generated via the \url{http://zooniverse.org} platform, development of which is funded by generous support, including a Global Impact Award from Google, and by a grant from the Alfred P. Sloan Foundation. The Digitised Sky Surveys were produced at the Space Telescope Science Institute under U.S. Government grant NAG W-2166. The images of these surveys are based on photographic data obtained using the Oschin Schmidt Telescope on Palomar Mountain and the UK Schmidt Telescope. The plates were processed into the present compressed digital form with the permission of these institutions. The Second Palomar Observatory Sky Survey (POSS-II) was made by the California Institute of Technology with funds from the National Science Foundation, the National Geographic Society, the Sloan Foundation, the Samuel Oschin Foundation, and the Eastman Kodak Corporation.

We implement our machine learning codes using Keras \citep{chollet2015} with a TensorFlow \citep{tensorflow2015} backend, and data augmentation using scikit-learn \citep{Pedregosa2011}. We also used Numpy \citep{oliphant2006}, Matplotlib \citep{hunter2007matplotlib} and Astropy  (\cite{astropy_collaboration2013}, \cite{astropy_collaboration2018}) Python3 \citep{python3} packages. Our codes are open source\footnote{https://github.com/matej-kosiba/CNN-multiwavelength\\-classification-of-X-ray-selected-galaxy-cluster-candidates}.

Matej Kosiba is supported by the \textit{European Space Agency} traineeship and the ERASMUS program and travel funding from the ESAC science faculty, Maggie Lieu was supported by \textit{European Space Agency} research fellowship at the European Space Astronomy Centre and a research fellowship at the University of Nottingham.




\bibliographystyle{mnras}
\bibliography{bibliography} 





\appendix
\section{Image preprocessing}
\label{Appendix:ImPreprocessing}

The output of the \xamin pipeline is an image with the following normalisation: 
if a pixel value is lower than min cut, it is attributed a value of 255; if a pixel is greater than max cut it is attributed a value of 0; and 255\,$\times$\,(1\,-\,(data\,-\,min cut)\,/\,(max~cut\,-\,min~cut)) otherwise \autoref{table_image_cuts}. To produce the .png images used in the neural networks, \xamin applies the normalisation separately to each of the channels according to \autoref{table_PNG_image_channel_values}.


\begin{table}
\caption{Threshold values used by the \xamin pipeline, std and median are the standard deviation and the median of the image data.}
\begin{center}
\begin{tabular}{c | c c } 
     & X-ray & Optical \\
    \hline
    min cut & 0 & median\,-\,std \\
    max cut & median\,$\times$\,14 & median\,+\,5\,$\times$\,std \\

\end{tabular}
\label{table_image_cuts}
\end{center}
\end{table}

\begin{table}
\caption{PNG image channel values as constructed by the \xamin pipeline. \textit{pix} refers to the pixel value after cutting.}
\begin{center}
\begin{tabular}{c | c c } 
    Channel & Pixel value &  Normalised pixel value  \\
    \hline
    \multirow{2}{*}{R} & pix\,>=\,176 & 255  \\
     & pix\,<\,176 & pix\,$\times$\,255\,/\,176  \\
    \hline
    \multirow{2}{*}{G} & pix\,>=\,120 & (pix\,-\,120)\,$\times$\,255\,/\,(255\,-\,120)  \\
     & pix.\,<\,120 & 0  \\
    \hline
    \multirow{2}{*}{B} & pix\,>=\,190 & (pix\,-\,190)\,$\times$\,255\,/\,(255\,-\,190)  \\
     & pix\,<\,190 & 0  \\

\end{tabular}
\label{table_PNG_image_channel_values}
\end{center}
\end{table}

\section{Extent -- extension likelihood plane plots}
\label{Appendix:Extent -- extent likelihood plots}
\begin{figure}
	\includegraphics[width=\columnwidth]{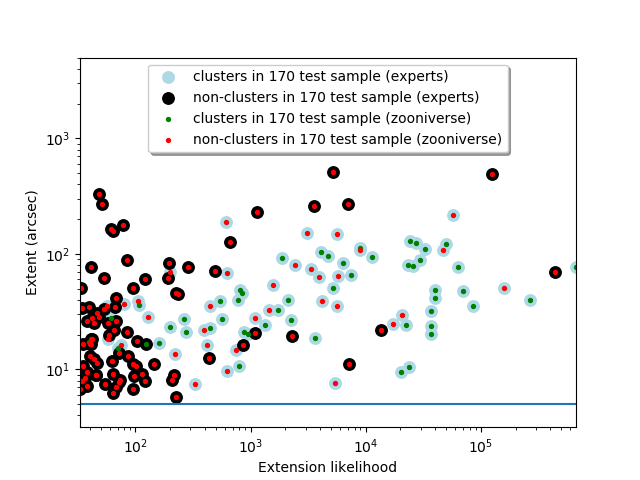}
	\caption{Extent -- extension likelihood plane for objects of the 170 test sample classified by experts and the Zooniverse volunteers.}
	\label{fig:EXT_EXTLIKE_experts_and_zooniverse_170_test_sample}
\end{figure}

\begin{figure}
	\includegraphics[width=\columnwidth]{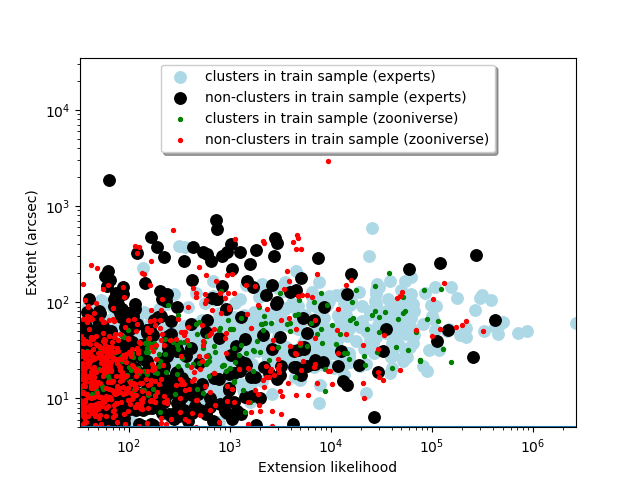}
	\caption{Extent -- extension likelihood plane for objects of the experts train sample and the Zooniverse train sample.}
	\label{fig:EXT_EXTLIKE_experts_and_zooniverse_train_sample}
\end{figure}

The extent -- extension likelihood plane plots (\autoref{fig:EXT_EXTLIKE_experts_and_zooniverse_170_test_sample}, \autoref{fig:EXT_EXTLIKE_experts_and_zooniverse_train_sample}) of our C1 sample of galaxy cluster candidates as described in \citep{Pacaud2006}, were used to analyse the Zooniverse sample of galaxy clusters and investigate our initial hypothesis, that the Zooniverse volunteers will preferentially find most prominent galaxy clusters. We find that the sample of the Zooniverse galaxy clusters span the entire extent -- extension likelihood plane and can not be recreated by a simple cut in this space. Please note however that the \xamin v3.5 we used to make the C1 sample had an issue fitting the point source peak, resulting in many non-clusters in the C1 region on the plots and that it is not the same pipeline as the XXL collaboration used before.  

\bsp	
\label{lastpage}
\end{document}